\documentstyle[epsf]{mn}

\def\eq#1 {eq.~(\ref{eq:#1})}

\def\Eq#1{Eq.~(\ref{eq:#1})}

\def\cl{{\it c.l.}}

\def\etal{{\it et al.\ }}

\def\cf{{cf.\ }}
\def\eg{{\it e.g.}}
\def\ie{{\it i.e.}}

\def\dnsig{$D_n-\sigma$}
\def\ltsima{$\; \buildrel < \over \sim \;$}
\def\lsim{\lower.5ex\hbox{\ltsima}}
\def\gtsima{$\; \buildrel > \over \sim \;$}
\def\gsim{\lower.5ex\hbox{\gtsima}}
\def\ga{\mathrel{\hbox{\rlap{\hbox{\lower4pt\hbox{$\sim$}}}\hbox{$>$}}}}
\def\la{\mathrel{\hbox{\rlap{\hbox{\lower4pt\hbox{$\sim$}}}\hbox{$<$}}}}

\def\kms{\,{\rm km\,s{^{-1}}}}
\def\hmpc{\,h{^{-1}}{\rm Mpc}}
\def\ihmpc{\,h{rm Mpc}{^{-1}}}

\def\ln{{\rm ln}}

\def\cos{{\rm cos}}
\def\sin{{\rm sin}}

\def\la{\langle} 
\def\ra{\rangle}

\def\iras{{\it IRAS}}

\def\pmb#1{\setbox0=\hbox{#1}%
 \kern-.025em\copy0\kern-\wd0
 \kern.05em\copy0\kern-\wd0
 \kern-.025em\raise.0433em\box0}

\def\vr{{\bf r}}
\def\br{{\bf r}}
\def\bv{{\bf v}}
\def\bm{{\bf m}}
\def\lcdm {$\Lambda$CDM} 
\def\ocdm {OCDM}

\def\hmpc{\, h^{-1} {\rm Mpc}}
\def\ihmpc{\, h\, {\rm Mpc^{-1}}}  
\def\3hmpc{\, ( h^{-1} {\rm Mpc})^3}
\def\kms{\, {\rm km\,s^{-1}}}  

\def\Pr{{\cal P}} 

\def\cL {{\cal L}}
\def\vv {{ \bf v}}

\def\bd {{ \bf d}}
\def\bx {{ \bf x}}

\def\WF {^{\rm WF}}
\def\CR {^{\rm CR}}

\def\rd{{\rm d}}

\def\Omeg {\Omega_0}

\def\r0p { r{_0^\prime}}
\def\prior {{\it prior}}

\def\uo{u^o}

\title{Large-Scale Power Spectrum and Structures From the ENEAR galaxy
Peculiar Velocity Catalog}

\author[Zaroubi, et al.]{S. Zaroubi$^1$,
        M. Bernardi$^{1,2,3}$,
        L.N. da Costa$^{2,4}$,
        Y. Hoffman$^5$,
        V. Alonso$^6$,
\newauthor        G. Wegner$^7$,
        C.N.A. Willmer$^{4,8}$,
        P.S. Pellegrini$^4$\\         
$^1$Max Planck Institut f\"ur Astrophysik, 
                 Karl-Schwarzschild-Str. 1,
                 85748 Garching, Germany.\\
$^2$European Southern Observatory, 
                 Karl-Schwarzschild-Str. 2,
                 85748 Garching, Germany.\\
$^3$Universitats-Sternwarte M\"unchen,
                  Scheiner-Str. 1, 
                  D-81679 M\"unchen, Germany.\\
$^4$Observat\'orio Nacional, 
                  Rua General Jose\'Cristino 77,
                  Rio de Janeiro, R. J. 20921 Brazil.\\
$^5$Racah Institute of Physics, The Hebrew University, Jerusalem 91904, Israel.\\
$^6$Observatorio Astron\'omico de C\'ordoba Laprida 854, C\'ordoba (5000), Argentina.\\
$^7$Department of Physics and Astronomy, 6127 Wilder
                  Laboratory, Dartmouth College, Hanover, NH
                  03755-3528.\\
$^8$UCO/Lick Observatory, University of California, 1156 High Street, Santa Cruz, CA 95064}

\begin{document}
\maketitle

\begin{abstract}

We estimate the {\it mass} density fluctuations power spectrum (PS) on
large scales by applying a maximum likelihood technique to the peculiar
velocity data of the recently completed redshift-distance survey of
early-type galaxies (hereafter ENEAR).  Parametric CDM-like models for
the PS are assumed, and the best fit parameters are determined by
maximizing the probability of the model given the measured peculiar
velocities of the galaxies, their distances and estimated errors.  The
method has been applied to CDM models with and without COBE
normalization. The general results are in agreement with the high
amplitude power spectra found from similar analysis of other independent
all-sky  catalogs of peculiar velocity data such as MARK III (Willick
\etal 1997)and SFI (Giovanelli \etal 1998; da Costa \etal 1996), 
in spite of the differences in the way these samples were selected, the
fact that they probe different regions of space and galaxy distances are
computed using different distance relations. For example, at $k=0.1
\ihmpc$ the power spectrum value is $P(k) \Omega^{1.2} =(6.5 \pm 3)
\times 10^3 (\hmpc)^3$ and $\eta_8\equiv\sigma_8 \Omega^{0.6} =
1.1_{-0.35}^{+0.2}$; the quoted uncertainties refer to $3\sigma$ error
level. We also find that, for \lcdm\ and \ocdm\ COBE-normalized models,
the best-fit parameters are confined by a contour approximately defined
by $\Omega h^{1.3}=0.377\pm0.08$ and $\Omega h^{0.88}=0.517\pm0.083$
respectively. $\Gamma$-shape models, free of COBE normalization, results
in the weak constraint of $\Gamma \geq 0.17$ and in the rather stringent
constraint of $\eta_8=1.0\pm 0.25$. All quoted uncertainties refer to
$3\sigma$ confidence-level (\cl).

The calculated PS has been used as a prior for Wiener reconstruction
of the density field at different resolutions and the
three-dimensional velocity field within a volume of radius $\approx 80
\hmpc$. All major structures in the nearby universe are recovered and
are well matched to those predicted from all-sky redshift surveys.
The robustness of these features has been tested with Constrained
Realizations (CR). Analysis of reconstructed three-dimensional
velocity field yields a small bulk flow amplitude ($\sim 160\pm
60~\kms$ at $60~\hmpc$) and a very small rms value of the tidal field
($\sim 60~\kms$). The results give further support to the picture that
most of the motion of the Local Group arises from mass fluctuations
within the volume considered.

\end{abstract}


\section{INTRODUCTION}
\label{sec:intro}


The canonical model of cosmology assumes that large-scale structure has
grown out of small density perturbations via the process of
gravitational instability.  These initial fluctuations are usually
assumed to satisfy the statistics of a Gaussian random field, solely
characterized by its power spectrum. In the linear regime, the
fluctuations grow self-similarly and retain their initial distribution
and power spectrum shape. Therefore, mapping the underlying cosmological
velocity field and its power spectrum on large scales, provides a direct
probe to the origin of structure in the universe.


The PS, the three-dimensional distribution of luminous matter and the
predicted peculiar velocity field have been derived from a variety of
data sets, especially from all-sky redshift surveys (\eg\ Strauss \&
Willick 1995 for a review of earlier work; Sutherland \etal 1999;
Branchini \etal 1999). Unfortunately, however, the distribution of
galaxies in these catalogs is not necessarily an unbiased tracer of
the underlying mass distribution, and suffer from the infamous
``galaxy biasing'' problem.  Furthermore, in estimates from redshift
surveys, uncertainties arise from the complicated relation between the
real space and the redshift space distributions, known as redshift
distortions (\eg , Kaiser 1987, Zaroubi and Hoffman 1996). In order to
avoid these problems altogether it is advantageous to appeal to
dynamical data, in particular catalogs of galaxy peculiar velocities
on large scales.


Peculiar velocities enable a direct and reliable determination of the
mass PS and distribution, under the natural assumption that the
galaxies are unbiased tracers of the large-scale,
gravitationally-induced, velocity field.  Furthermore, since peculiar
velocities are non-local and have contributions from different scales,
analysis of the peculiar velocity field provides information on scales
somewhat larger than the sampled region (\eg\, Hoffman \etal 2000).
For the same reason peculiar velocities are adequately described by
linear theory even when densities become quasi-linear (\eg , Freudling
\etal 1999). Consequently, the dynamics and the distribution of
peculiar velocities are well described by the linear regime of
gravitational instability and by a Gaussian probability distribution
function (PDF), respectively.


Assuming that both the underlying velocity field and the errors are
drawn from independent random Gaussian fields, the observed peculiar
velocities constitute a multi-variant Gaussian data set, albeit the
sparse and inhomogeneous sampling.  The corresponding {\it posterior}
PDF is a multivariate Gaussian that is completely determined by the
assumed PS and covariance matrix of errors. Under these conditions one
can write the joint PDF of the model PS and the underlying velocity or
density field.  


The purpose of the present study is to calculate, from the joint PDF,
the PS and 3D mass distribution, as well as the 3D peculiar velocity
field, as derived from the newly completed ENEAR galaxy peculiar
velocity catalog (da Costa \etal 2000a, Paper~I). First, the PS model
parameters are estimated by maximizing the likelihood function given
the model (Zaroubi \etal 1997). An identical likelihood estimation of
the power spectrum has been previously applied to the {\rm Mark III} (Zaroubi
\etal 1997) and the SFI (Freudling \etal 1999) data sets. In both
cases the analysis yielded a high amplitude power spectrum. Although
the results from those two catalogs are consistent with each other,
they are marginally inconsistent with the power spectra measured from
redshift catalogs (\eg , da Costa \etal 1996; Sutherland \etal 1999),
inferred from the analysis of the velocity correlation function (\eg\,
Borgani \etal 2000a, 2000b), and from velocity-velocity comparisons (\eg
, Davis \etal 1996, da Costa \etal 1998). One of our goals is to use
the same methodology employed before for the Mark~III and SFI to the
new ENEAR catalog to directly test the reproducibility of the results
with an independent sample based on a different distance indicator but
probing a comparable volume.




Second, the Wiener filter (WF) solution of the field is recovered by
finding the most probable field given the PS and the data (Zaroubi \etal
1995, 1999). Constrained realizations (CR) are then used to sample the
statistical scatter around the WF field (Hoffman and Ribak 1991). The
mass density PS is used to calculate the smoothed Wiener filtered
density and 3D velocity fields given the measured radial velocities
(Zaroubi \etal 1995, 1999).  The WF provides an optimal estimator of the
underlying field in the sense of a minimum-variance solution given the
data and an assumed \prior\ model (Wiener 1949; Press \etal\ 1992). The
\prior\ defines the data auto-correlation and the data-field
cross-correlation matrices.  In the case where the data is drawn from a
random Gaussian field, the WF estimator coincides with the conditional
mean field and with the most probable configuration given the data (see 
Zaroubi \etal 1995).  It should be noted that Kaiser \& Stebbins (1991) 
were the first to propose a Bayesian solution to the problem of 
reconstruction from peculiar velocity data sets.

Finally, the recovered three-dimensional velocity field is used to
compute the amplitude of the bulk flow and to decompose the velocity
field in terms of a divergent and tidal components which enables one
to separate the contribution to the measured peculiar velocity field
from mass fluctuations within and outside the volume probed by the
data (Hoffman \etal 2000).


The methods adopted in this study do not involve any explicit window
function, weighting or smoothing the data.  In addition, they
automatically underweight noisy, unreliable data.  However, a few
simplifying assumptions are required: 1) peculiar velocities are drawn
from a Gaussian random field; 2) peculiar velocities are related to the
densities through linear theory; 3) errors in the $D_n-\sigma$ inferred
distances constitute a Gaussian random field with two components, the
first scales linearly with distance while the second models the
nonlinear evolution of the velocities as a constant scatter.  The need
to assume a parametric functional form for the PS is also a limitation.


The outline of this paper is as follows. In \S~\ref{sec:data} we
briefly describe the peculiar velocity data used in the present
analysis.  The PS analysis is carried out in \S~\ref{sec:PS}.  The
Wiener filtering is applied to the ENEAR data in \S~\ref{sec:wiener},
where maps of the density field are presented and compared to those
predicted from redshift surveys. Also shown in this section are the
recovered three-dimensional velocity field and the results of its
analysis. Our results are summarized and discussed in
\S~\ref{sec:conclusion}.


\section{The Data}
\label{sec:data}

In the present analysis, we use the ENEAR redshift-distance survey
described in greater detail in Paper~I of this series. Briefly, the
ENEAR sample consists of roughly 1600 early-type galaxies brighter
than $m_B=14.5$ and with $cz \leq 7000~\kms$, for which \dnsig
distances are available for 1359 galaxies. Of these 1145 were deemed
suitable for peculiar velocity analysis according to well-defined
criteria (Paper~I; M. Alonso \etal, in preparation). To the
magnitude-limited sample we added 285 fainter and/or with redshifts $>
7000~\kms$, 129 within the same volume as the magnitude-limited
sample. In total, the cluster sample consists of 569 galaxies in 28
clusters, which are used to derive the distance relation (M. Bernardi
\etal, in preparation).  Over 80\% of the galaxies in the
magnitude-limited sample and roughly 60\% of the cluster galaxies have
new spectroscopic and R-band photometric data obtained as part of this
program. Furthermore, repeated observations of several galaxies in the
sample (M. Alonso \etal, in preparation; G. Wegner \etal 2000) provide
overlaps between observations conducted with different
telescope/instrument configurations and with data available from other
authors. These overlaps are used to tie all measurements into a common
system, thereby ensuring the homogeneity of the entire dataset. In
contrast to other samples new observations conducted by the same group
are available over the entire sky.  The comparison between the sample
of galaxies with distances and the parent catalog also shows that the
sampling across the sky is uniform.

Individual galaxy distances were estimated from a direct \dnsig\
template relation derived by combining all the available cluster data
(M. Bernardi \etal, in preparation), corrected for incompleteness and
associated diameter-bias (Lynden-Bell \etal 1988). From the observed
scatter of the template relation the estimated fractional error in the
inferred distance of a galaxy is $\Delta \sim 0.19$, nearly
independent of the velocity dispersion.

Since early-type galaxies are found preferentially in high-density
regions, galaxies have been assigned to groups/clusters using
well-defined criteria imposed on their projected separation and
velocity difference relative to the center of groups and
clusters. These systems were identified using objective algorithms
applied to the available magnitude-limited samples, comprising all
morphological types, with complete redshift information probing the
same volume. For membership assignment we used group catalogs
published by Geller \& Huchra (1983), Maia, da Costa \& Latham (1988),
Ramella, Pisani \& Geller (1997) as well as unpublished results
(M. Ramella \etal, in preparation) covering other regions of the sky.
The characteristic size and velocity dispersion of these
groups/clusters were used to establish the membership of the ENEAR
early-types, as described in Paper~I. We find isolated galaxies,
groups with only one early-type, and groups with two or more
early-types. Early-type galaxies in a group/cluster are replaced by a
single object having: (1) the redshift given by the group's mean
redshift, which is determined considering all morphologies; (2) the
distance given by the error-weighted mean of the inferred distances,
for groups with two or more early-types; and (3) the fractional
distance error given by $\Delta / \sqrt(N)$, where $N$ is the number
of early-types in the group. In some cases groups were identified with
Abell/ACO clusters within the same volume as the ENEAR sample and
fainter cluster galaxies were added, as described in Paper~I. In the
analysis below we compute the dipole component of the velocity field
out to $6000~\kms$ as probed by all objects, and by splitting the sample
into two independent sub-samples consisting of field galaxies and
groups/clusters. The latter is done to evaluate directly from data the
amplitude of possible sampling errors.

The inferred distances are corrected for the homogeneous and
inhomogeneous Malmquist bias (IMB). The latter was estimated using the
PCSz density field (Branchini \etal 1999), corrected for the effects
of peculiar velocities, in the expressions given by Willick \etal
(1997). In this caculation we also include the correction for the
redshift limit of the sample. A complete account of the sample used
and the corrections applied will be presented in a subsequent paper of
this series (M. Alonso \etal, in preparation).

\section{Power Spectrum}
\label{sec:PS}

The calculation of the matter PS from the peculiar velocity data by
means of likelihood analysis requires a relation between the velocity
correlation function and the power spectrum. Define the two-point
velocity correlation ($3\times 3$) tensor by the average over all pairs
of points $\vr_i$ and $\vr_j$ that are separated by $\vr=\vr_j-\vr_i$,
\begin{equation}\Psi_{\mu\nu} (\vr) \equiv \la v_\mu (\vr_i) v_\nu (\vr_j) \ra,
\label{eq:vcorr} \end{equation}where $v_\mu(\vr_i)$ is the $\mu$ component of the
peculiar velocity at $\vr_i$. In linear theory, it can be expressed in
terms of two scalar functions of $r=\vert \vr \vert$ (G\'orski 1988),
computed from the parallel and perpendicular components of the
peculiar velocity, relative to the separation vector $\vr $, \begin{equation}
\Psi_{\mu\nu}(\vr) = \Psi_{\perp}(r) \delta_{\mu\nu} +
[\Psi_{\Vert}(r) - \Psi_{\perp}(r)] \hat \br_\mu \hat \br_\nu. \end{equation}The
spectral representation of these radial correlation functions is \begin{equation}
\Psi_{\perp,\Vert}(r)= {H_0^2 f^2(\Omega)\over 2 \pi^2} \int_0^\infty
P(k)\, K_{\perp,\Vert}(kr)\, dk , \label{eq:Psi} \end{equation}where
$K_{\perp}(x) = j_1(x)/ x$ and $K_{\Vert}(x) = j_0-2{j_1(x)/ x}$, with
$j_l(x)$ the spherical Bessel function of order {\it l}. The
cosmological $\Omega$ dependence enters as usual in linear theory via
$f(\Omega)\approx \Omega^{0.6}$, and $H_0$ is the Hubble constant. A
parametric functional form of $P(k)$ thus translates to a parametric
form of $\Psi_{\mu\nu}$. Note that the quantity that can be derived
from peculiar-velocity data via the linear approximation is
$f^2(\Omega)\, P(k)$, where $P$ is the mass density PS.

Let $\bm$ be the vector of model parameters and $\bd$ the vector of $N$
data points. Then Bayes' theorem states that the {\it posterior}
probability density of a model given the data is

\begin{equation}\Pr (\bm \vert \bd ) = {\Pr(\bm) \Pr(\bd|\bm) \over \Pr(\bd)}. \end{equation}

The denominator is merely a normalization constant.  The probability
density of the model parameters, $\Pr(\bm)$, is unknown, and in the
absence of any other information we assume it is uniform within a
certain range.  The conditional probability of the data given the
model, $\Pr(\bd|\bm)$, is the likelihood function, ${\cal
L}(\bd|\bm)$.  The objective in this approach, which is to find the
set of parameters that maximizes the probability of the model given
the data, is thus equivalent to maximizing the likelihood of the data
given the model (\cf Zaroubi \etal 1997; Jaffe \& Kaiser 1994).  The
Bayesian analysis measures the relative likelihood of different
models. An absolute frequentist's measure of goodness of fit is
provided by the $\chi^2$ per degree of freedom (hereafter, {\it
d.o.f.}), which we use as a check of the best parameters obtained by
the likelihood analysis.

Assuming that the velocities are a Gaussian random field,
the two-point velocity correlation tensor $\Psi$ fully characterizes the
statistics of the velocity field.
Define the radial-velocity correlation ($N\times N$) matrix $U_{ij}$ by
\begin{equation}
U_{ij} = \hat \br_i\dag\, \Psi \,\hat\br_j=
\Psi_{\perp}(r)\sin\theta_i \sin\theta_j + \Psi_{\Vert}(r)\cos\theta_i
\cos\theta_j  \, ,
\label{eq:ucorr}
\end{equation}
where $i$ and $j$ refer to the data points, 
$r=\vert \br \vert=\vert \br_j-\br_i \vert$ and the angles are
defined by $\theta_i=\hat{\br_i}\cdot\hat{\br}$ (G\'orski 1988; Groth,
Juszkiewicz \& Ostriker 1989).
Let the inferred radial peculiar velocity at $\vr_i$ be $\uo_i$,
with the corresponding error $\epsilon_i$ also assumed to be a
Gaussian random variable.  The observed correlation matrix is then
$\tilde U_{ij} = U_{ij} + \epsilon_i^2 \delta_{ij}$, and
the likelihood of the $N$ data points is
\begin{equation}
{\cal L} = [ (2\pi)^N \det(\tilde U_{ij})]^{-1/2}
  \exp\left( -{1\over 2}\sum_{i,j}^N {\uo_i \tilde U_{ij}^{-1} \uo_j}\right) .
\label{eq:like}
\end{equation}

Given that the correlation matrix, $\tilde U_{ij}$, is symmetric and
positive definite, we can use the Cholesky decomposition method
(Press \etal 1992) for computing the likelihood function (\Eq{like}).
The significant contribution of the errors to the diagonal terms
makes the matrix especially well-conditioned for decomposition.


The errors are assumed to have two contributions, the first is the
usual $D_n-\sigma$ distance proportional errors (about $19\%$ per
galaxy for ENEAR). The second is a constant error that accounts for
the non-linear velocities of galaxies in the high density environment
in which early-type galaxies reside. This term represents our poor
understanding of the complex correlations introduced by non-linear
evolution. For each power spectrum model, we have performed the
likelihood analysis assuming this constant value to be either null or
$ 250 \kms$ but, as shown below, the difference in the results are
only marginal and do not affect our general conclusions.

\subsection{COBE-Normalized CDM Models}
\begin{figure}
\setlength{\unitlength}{1cm} \centering
\begin{picture}(9,16)
\put(-1, 8.5){\includegraphics{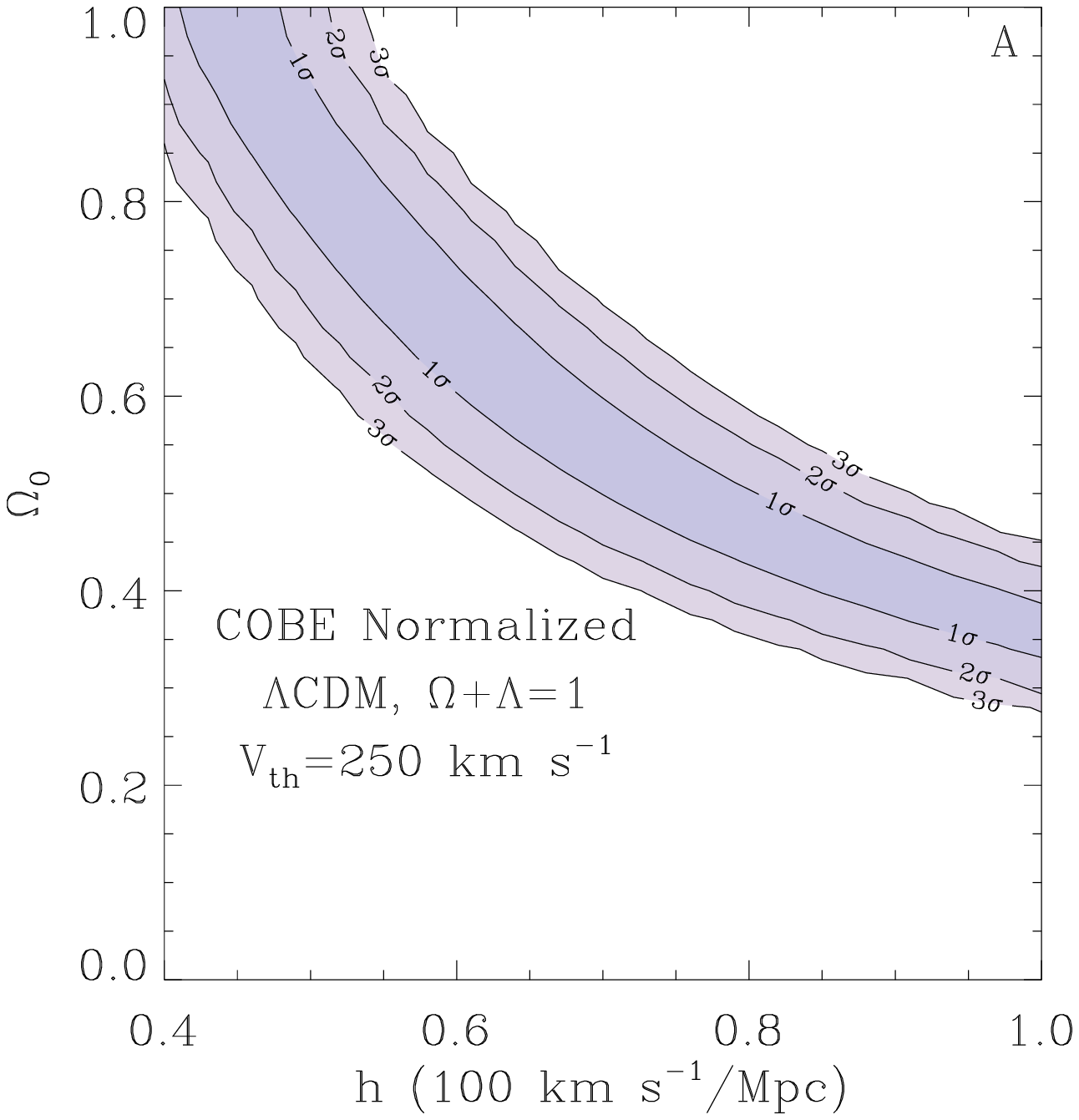}}
\put(-1, 0){\includegraphics{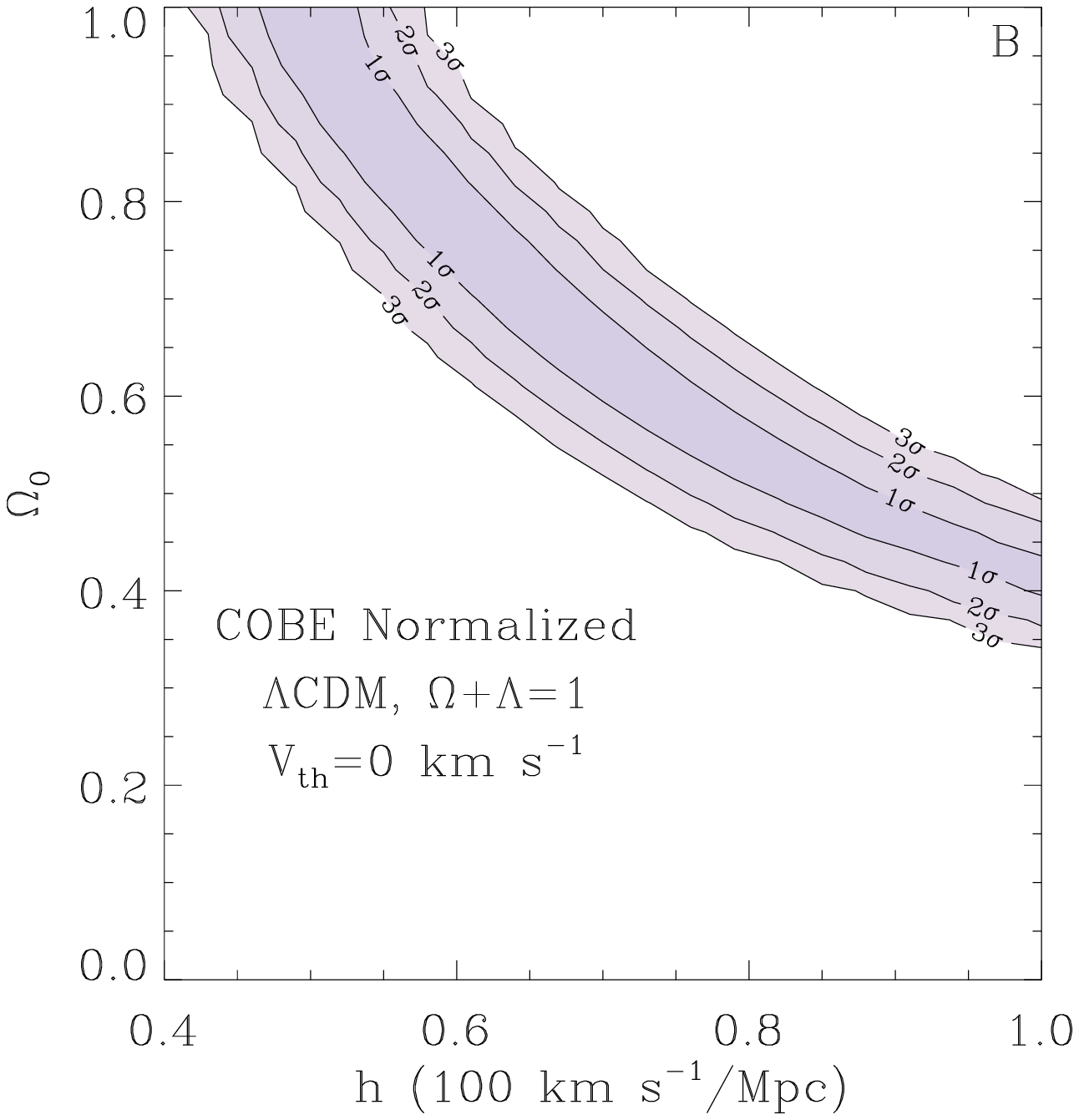}}
\end{picture}
\caption{ Contour map of ln-likelihhod in the $h -\Omega$ plane for
\lcdm\ $\;$ models with $250 \kms$ thermal error component (upper
panel) and with zero thermal error (lower panel). The contours
denote the most likely values within $1,2$ and $3\sigma$ \cl.}
\label{fig:fig1}
\end{figure}

We first restrict our attention to the generalized family of CDM
cosmological models, allowing variations in the cosmological
parameters $\Omega$, $\Lambda$ and $h$.  Furthermore, four-year COBE
normalization is imposed as an additional external constraint.  The
general form of the PS for these models is 
\begin{equation}P(k) = A_{\rm
COBE}(n,\Omega,\Lambda)\, T^2(\Omega,\Omega_B,h; k)\, k^n, \end{equation}where
the CDM transfer function proposed by Sugiyama (1995) is adopted,

\begin{eqnarray}
T(k)& = & {\ln\left(1+2.3q) \right)\over 2.34q}\nonumber \\ &  &
\left[1+3.89q+(16.1q)^2+(5.46q)^3+(6.71q)^4\right]^{-1/4} \quad,
\label{eq:transfunc}
\end{eqnarray}
\begin{equation} q=k \left[ \Omega h\, \exp (-\Omega_b -h_{50}^{1/2}
\Omega_b/\Omega)\, (\ihmpc) \right]^{-1}.  \end{equation}
The parameters $\Omega$ and $h$ are varied such that they span the
range of currently popular CDM models, including \lcdm\
($\Omega+\Lambda=1$, $\Omega\leq 1$) and \ocdm\ ($\Lambda=0$,
$\Omega\leq 1$).  In all cases, the baryonic density is assumed to be
$\Omega_b =0.019 h^{-2}$, which is the value currently favored by
primordial nucleosynthesis analysis (\eg , Burles \& Tytler 1998). We
limit our inquiry to models without tilt, namely to models with
$n=1$. For each model, the normalization of the PS is fixed by the
COBE 4-year data (Bennet \etal 1996); for more details see Zaroubi
\etal (1997 and references therein).

Figure 1a shows the likelihood contour map in the $\Omega-h$ plane,
for the \lcdm\ family of models with $n=1$ (normalization by Sugiyama
1995), the error matrix is assumed here to have an additional
diagonal random contribution of $250 \kms$ that accounts for the
nonlinear evolution of the galaxies.
The most probable parameters in this case (in the range $\Omega\leq1$)
are $\Omega=1$ and $h=0.5$. The elongated contours clearly indicate
that neither $\Omega$ nor $h$ are independently well constrained.
It is rather a degenerate combination of the two parameters, approximately
$\Omega h^x$ with $x\sim 1$ (\ie\ a combination
close to the $\Gamma$ parameter) that is being determined tightly
by the elongated ridge of high likelihood.

Figure 1b shows the likelihood results for the same \lcdm\ model shown
in Fig. 1a but with no random contribution to the error matrix.  The
contours in Fig. 1b show very little changes relative to those shown
in panel (a), notably they get tighter and the best values of $\Omega$
for a given Hubble constant are somewhat higher. The addition of a
reasonable random component to the error matrix does not alter the
results in any significant way for any of the PS models considered in
this study. For the rest of the PS models we show the calculation with
the addition of a constant error of $250 \kms$.

Figure 2 shows the similar likelihood map for \ocdm\ with $n=1$.
The most probable values here are $\Omega=0.53$ and $h=1$. The values
of $\Omega$ and $h$ are not independently constrained here as well.

We can thus quote stringent constraints on the conditional best value of
$\Omega$ given $h$ for the COBE normalized CDM models shown in Figs. 1a
and 2: $\Omega\approx (0.377\pm 0.08) h^{-1.3}$ for \lcdm, and
$\Omega\approx (0.517\pm 0.083) h^{-0.88}$ for \ocdm.

\begin{figure}
\setlength{\unitlength}{1cm} \centering
\begin{picture}(8,8)
\put(-2,-3){\includegraphics{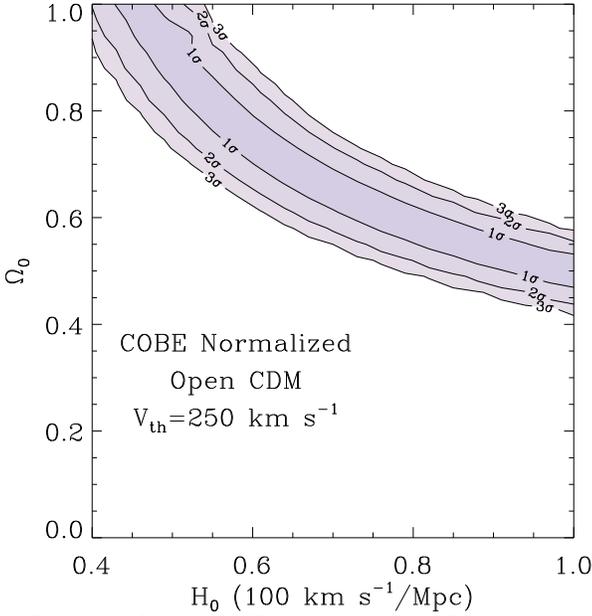}}
\end{picture}
\caption{Same as in Figure 1a but for \ocdm\ models.}
\label{fig:fig2}
\end{figure}

\subsection {The $\Gamma$ Model}
\begin{figure}
\setlength{\unitlength}{1cm} \centering
\begin{picture}(8,8)
\put(-2, -3){\includegraphics{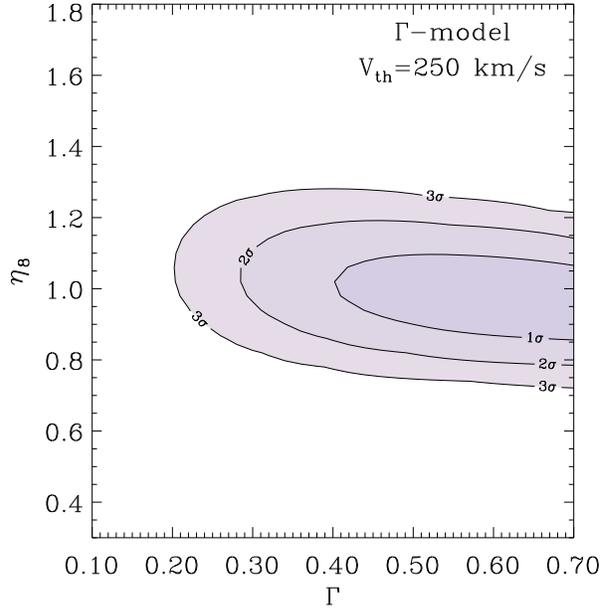}}
\end{picture}
\caption{Contour map of ln-likelihood for the $\Gamma$ model in the
$\Gamma-\eta8$ plane. The contours denote the $1, 2$ and $3\sigma$}
\cl.\label{fig:fig3}
\end{figure}

To recover the PS from the velocity data independent of the COBE normalization,
we use as a parametric prior the so-called $\Gamma$ model
(\eg , Efstathiou, Bond and White 1992),
\begin{eqnarray}
P(k) & = & A\, k\, T^2(k), \nonumber \\ 
T(k) & = & \Bigl( 1 + [ ak/\Gamma + (bk/\Gamma)^{3/2} + (ck/\Gamma)^2 ] ^{\nu}
\Bigr)^{-1/\nu} \quad,
\label{eq:gamma}
\end{eqnarray}
with $a=6.4\hmpc$, $b=3.0\hmpc$, $c=1.7\hmpc$ and $\nu = 1.13$. The free
parameters to be determined by the likelihood analysis are the
normalization factor $\eta_8 \equiv\sigma_8 \Omega^{0.6}$ and the 
$\Gamma$ parameter. In the context of the CDM cosmological model,
$\Gamma$ has a specific cosmological interpretation, $\Gamma=\Omega h$.
Here, however, \Eq{gamma} serves as a generic function with logarithmic
slopes $n=1$ and $-3$ on large and small scales respectively, and with a
turnover at some intermediate wavenumber that is determined by the
single shape parameter $\Gamma$.

Figure 3 shows the contour map of $\ln \cL$ in the
$\Gamma\,$--$\,\eta_8$ plane.  Although the likelihood analysis poses
strong constraint on the allowed values of $\eta_8$
($=1._{-0.28}^{+0.3}$ with $3\sigma$ \cl), it only weakly constrains
the value of $\Gamma$ ($\ge 0.18$ with $3\sigma$ \cl), and
$\Gamma=0.25$ is excluded with $2\sigma$ \cl.

\subsection{Results and Comparison between the Various Models}

The best fit models for each CDM family have a comparable likelihood, 
with the most likely model is the \ocdm\ model with $\Omega=0.53$ and
$h=1$. All best fit models agree within $\approx 20\%$ for $k > 0.1
\ihmpc$. The amplitude of the PS at $k=0.1 \ihmpc$ for all models lies
within $P(k) \Omega^{1.2} =(6.5 \pm 3) \times 10^3 (\hmpc)^3$ and the
values of $\eta_8$ are within the range $1.1_{-0.35}^{+0.2}$.

Figure 4 shows the power spectrum of the most-likely COBE-normalized
model and the $3\sigma$ errors about it. It also shows the PS
corresponding to most likely models of \lcdm\ and $\Gamma$ models.
Within the errors, the most likely PS for each CDM family are very
consistent, especially at intermediate scales ($30-50\hmpc$), where the
data information content provide the strongest constraint. Also shown in
Fig. 4 are the best-fit PS, obtained from a similar likelihood analysis,
for the {\rm Mark III} and SFI data sets. As can be seen the most likely PS for the
three catalogs are in good agreement. This result shows that the high
amplitude PS found from peculiar velocity data is unlikely to be due to
possible non-uniformities of these catalogs or to the type of galaxies
used. In fact, while  Mark~III and SFI relied predominantly on TF
distances to spirals, ENEAR relies on $D_n-\sigma$ distances to
early-type galaxies. On the other hand, the reason for the discrepancy
in the cosmological constraints between the maximum likelihood method 
and other methods (da Costa \etal 1998; Strauss \& Willick 1998; Borgani
\etal 2000a,2000b) remains unresolved. The former yields a systematically
higher amplitude PS, as reflected by the high values of $\eta_8$, which
is also  in disagreement with the constraints derived from other
analysis of LSS data.  Possible explanations are given in
\S~\ref{sec:conclusion}.

In all the COBE-normalized PS models considered the $\chi^2/d.o.f.$ of
the best fit models is of the order of 0.93. This value deviates by
about $2\sigma$ from the $\chi^2/d.o.f$ desired value of unity. This,
however, does not pose any serious problem since many of the models
within the likelihood most likely contours have a $\chi^2/d.o.f.\simeq
1$. The $\chi^2/d.o.f$ for the $\Gamma$-model is 0.99.

\begin{figure}
\setlength{\unitlength}{1cm} \centering
\begin{picture}(7,8)
\put(-2.5, -2.8){\includegraphics{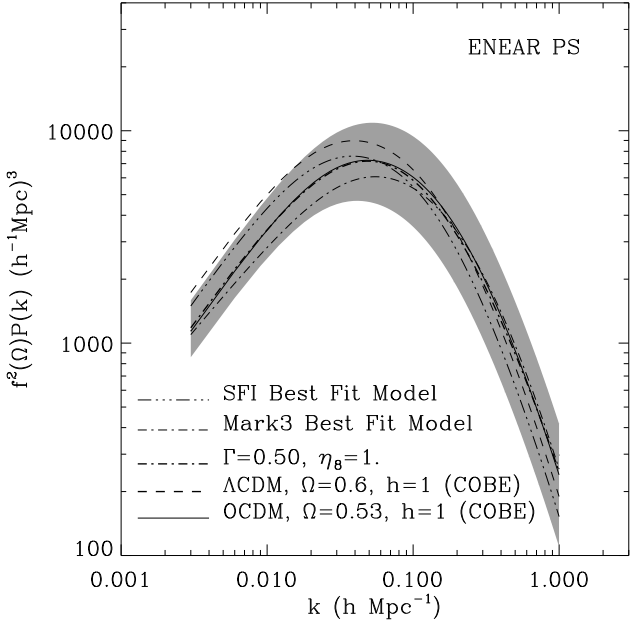}}
\end{picture}
\caption{The PS of the most probable COBE-normalized \ocdm\ 
(solid-bold) and \lcdm\ (dashed-bold) models and of the $\Gamma$-model (dotted-dashed-bold).
shown also the most probable models as estimated from {\rm Mark III} (dotted-dashed)
and SFI (triple-dotted-dashed) data sets. The shaded region around the PS marks the 3-$\sigma$ 
\cl\, note that the dynamical range of the data is confined to $0.05\lsim k \lsim 0.3$.}
\label{fig:fig4}
\end{figure}

\section{Wiener Filter \& Constrained Realizations}
\label{sec:wiener}

\subsection{The Method}

Having determined the power spectrum, all the ingredient needed to
Wiener reconstruct the density and velocity fields are ready.  Details
on the general application of the WF/CR method to the reconstruction of
large-scale structure are described in Zaroubi \etal (1995), where the
theoretical foundation is discussed in relation with other methods of
estimation, such as Maximum Entropy. The specific application of the
WF/CR method to peculiar velocity data  sets has been presented in
Zaroubi \etal (1999). Here we provide only a brief description of the
WF/CR method, for more details the reader is referred to the original
references references given above.

We assume that the peculiar velocity field $\vv(\vr)$
and the density fluctuation
field $\delta(\vr)$ are related via linear gravitational-instability theory.
Under the assumption of a specific theoretical prior for the power
spectrum $P(k)$ of the underlying density field,
one can write the WF minimum-variance estimator of the fields as
\begin{equation}
\bv\WF(\br) = \Bigl < \bv(\br) \uo_i \Bigr >  \Bigl < \uo_i \uo_j \Bigr >
^{-1}        \uo_j
\label{eq:WFv}
\end{equation}
and
\begin{equation}
\delta\WF(\br) = \Bigl < \delta(\br) \uo_i \Bigr >  \Bigl < \uo_i \uo_j
\Bigr > ^{-1}    \uo_j  .
\label{eq:WFd}
\end{equation}

A well known problem of the WF is that it attenuates the estimator to
zero in regions where the noise dominates. The reconstructed mean
field is thus statistically inhomogeneous.  In order to recover
statistical homogeneity we produce constrained realizations (CR), in
which random realizations of the residual from the mean are generated
such that they are statistically consistent both with the data and the
\prior\ model (Hoffman and Ribak 1991; see also Bertschinger 1987).
In regions dominated by good quality data, the CRs are dominated by
the data, while in the limit of no data the realizations are
practically unconstrained.

The CR method is based on creating random
realizations, $\tilde\delta(\br)$ and $\tilde\bv(\br)$,
of the underlying fields that obey the assumed PS and
linear theory, and a proper set of random errors $\tilde\epsilon_i$.
The velocity random realization is then ``observed" like the actual data to
yield a mock velocity dataset $\tilde \uo_i$.  Constrained realizations
of the dynamical fields are then obtained by
\begin{equation}
\bv \CR(\br) = \tilde \bv(\br) + \Bigl < \bv(\br) \uo_i \Bigr >
\Bigl < \uo_i \uo_j \Bigr > ^{-1} \bigl( \uo_j -\tilde \uo_j \bigr)
\label{eq:CRv}
\end{equation}
and
\begin{equation}
\delta \CR(\br) =
\tilde \delta(\br) +  \Bigl < \delta(\br) \uo_i \Bigr >  \Bigl < \uo_i \uo_j
\Bigr > ^{-1} \bigl( \uo_j -\tilde \uo_j \bigr) .
\label{eq:CRd}
\end{equation}
The two types of covariance matrices in the above equations are computed
within the framework of linear theory as follows.
The covariance matrix of the data $\Bigl < \uo_i \uo_j \Bigr >$ is the same matrix 
$\tilde U_{ij}$ that appears in eq.~\ref{eq:like}.

The cross-correlation matrix of the data and the underlying field
enters the above equations as, \eg ,
\begin{equation}
\Bigl < \delta(\br) \uo_j \Bigr >
=\Bigl < \delta(\br) \bv(\br_j) \Bigr > \cdot \hat \br_j .
\end{equation}
The two-point cross-correlation vector between the density and velocity fields
is related to the PS via
\begin{equation}
\Bigl < \delta(\bx)\, \bv(\bx + \br) \Bigr > =
- {H_0 f(\Omeg)\over 2  \pi^2} \hat\br
\int_0^\infty k P(k) j_1(kr) \rd k.
\label{eq:cross}
\end{equation}

The assumption that linear theory is valid on all scales enables us to
choose the resolution as well, and in particular to use different
smoothing radii for the data and for the recovered fields.
In our case no smoothing were applied to the radial velocity data
while we choose to reconstruct the density field with a finite
Gaussian smoothing of radius $R$. This alters the density-velocity
correlation function by inserting the multiplicative term
$\exp[-k^2 R^2/2] $ into the integrand of \eq{cross} .

A theoretical estimate of the signal-to-noise ratio ($S/N$)  at every
point in space is given by a simple expression (see  Zaroubi \etal 1999)
but it requires the calculation and inversion of very  large matrices.
Therefore,  in this study we estimate the point to  point error by
conducting a large number of CRs. In the case of random Gaussian fields,
the ensemble of CRs defined in \eq{CRv} and \eq{CRd}  samples the
distribution of uncertainties in the mean Wiener density  and velocity
fields (Hoffman \& Ribak 1991).

It is worth noting that the WF represents a general minimum-variance
solution under the sole assumption that the field is a random field
with a known power spectrum.  No assumption has to be made here
regarding higher order correlations (or the full joint probability
distribution functions) of the underlying field.  On the other hand,
the CRs are derived under the explicit assumption of a full Gaussian
random field.

\subsection{Maps of Density and Velocity Fields}
\begin{figure}
\setlength{\unitlength}{1cm} \centering
\begin{picture}(7,9)
\put(-3., -2){\includegraphics{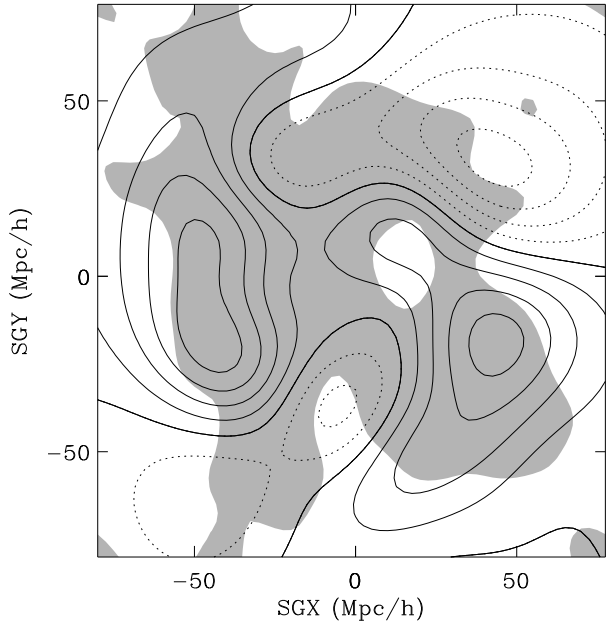}}
\end{picture}
\caption{Reconstruction from the ENEAR catalog. Left panel: The
density field map in the Supergalactic plane, with G12 smoothing. 
Density contour spacing is 0.1, positive contours are solid, negative
contours are dashed and $\delta=0$ is denoted by heavy-solid line. 
The shading indicates regions where the error is less than $0.3$.}
\label{fig:fig5}
\end{figure}

Figure 5 shows the map of the density field along the Supergalactic
plane obtained from the ENEAR data using a Gaussian smoothing radius
of $1200 \kms$ (hereafter G12). The shaded area corresponds to the
region where the error, as estimated from performing 10 CRs, in
density is less than 0.3. The main features of our local universe are
easily identified in the WF map, including the Great Attractor (GA) on
the left and the Perseus-Pisces supercluster (PP) in the lower
right. There is also a hint of the Coma cluster, which lies just
outside the sample, in the upper part on the map. Even though
different in details, the gross features of the density field are
remarkably similar to those obtained by Zaroubi, Hoffman \& Dekel
(1999) from the application of the same formalism to the Mark~III
catalog. This is an outstanding result considering the different ways
the two catalogs were constructed and the pecuiar velocities measured.

\begin{figure}
\setlength{\unitlength}{1cm} \centering
\begin{picture}(8,16)
\put(-2.5, 6){\includegraphics{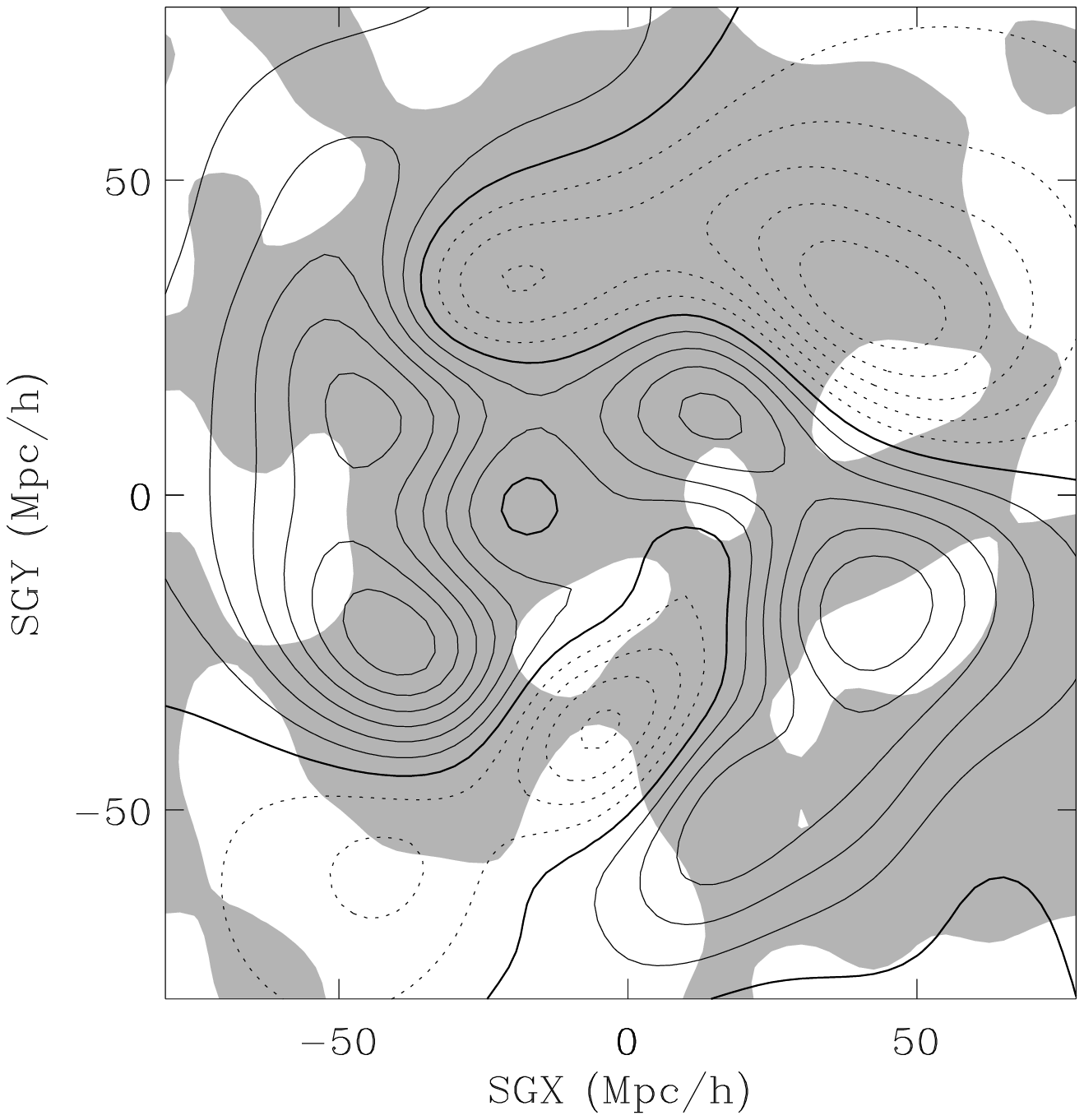}}
\put(-2.5, -2.5){\includegraphics{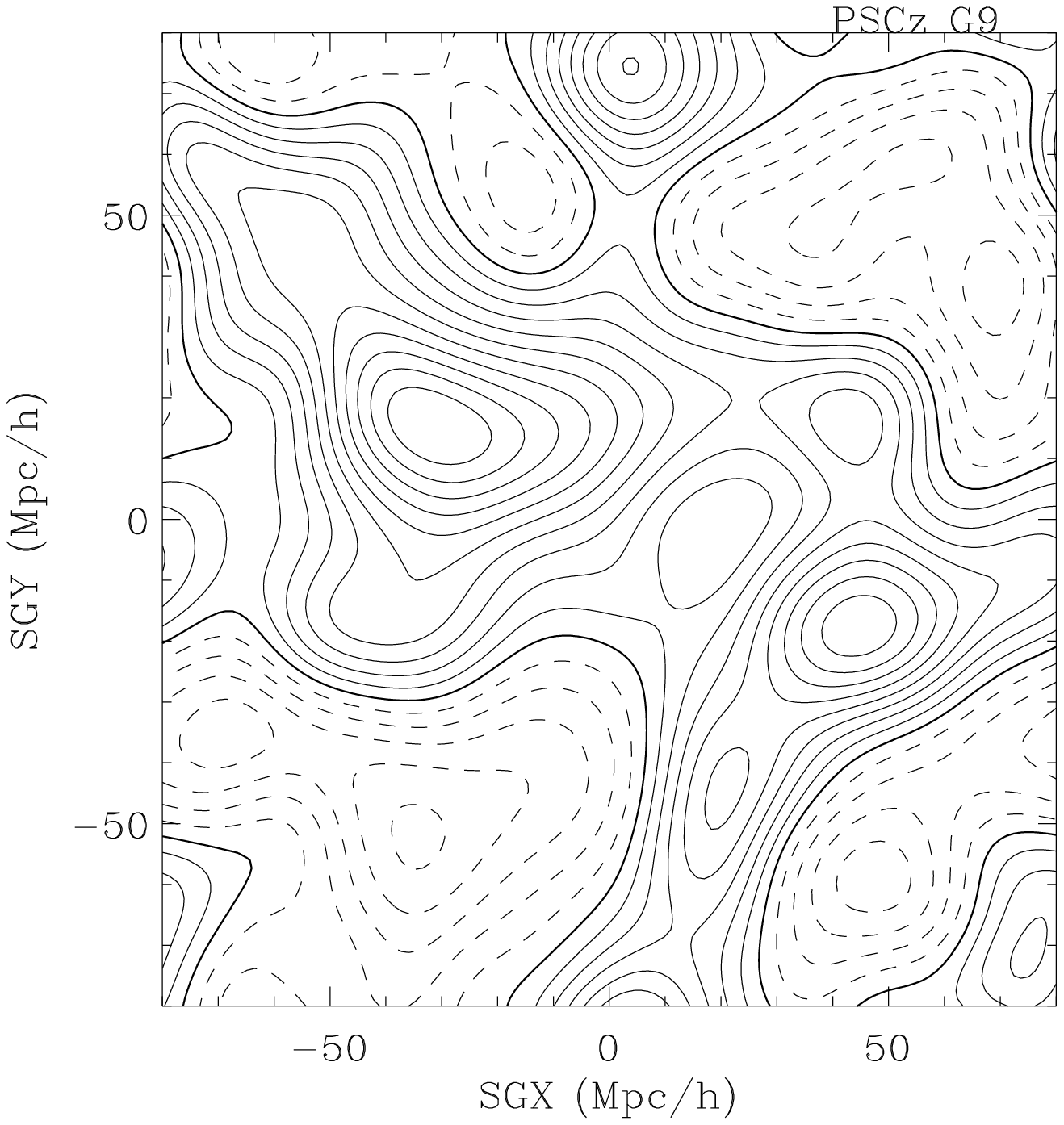}}
\end{picture}
\caption{Upper panel: The same as in Fig. 5 but with G9 smoothing. The shaded
area indicates regions with error smaller than $0.45$. Lower panel:
The G9 smoothing density reconstruction from the PSCz redhsift catalog
(Branchini \etal 1999)}
\label{fig:fig6}
\end{figure}

Fig. 6 compares a higher resolution map of the density field recovered
from the ENEAR data (left panel) to the density field reconstructed
from the PSCz redshift catalog (right panel; Branchini \etal
1999). Both maps are along the Supergalactic plane and were
reconstructed using a $900 \kms$ smoothing radius. The shaded area in
the left panel indicates regions where the error is less than
0.45. Even though different in detail the similarities between the
density fields are striking lending further credence to the reality of
the structures observed in the mass distribution. Note that with the
higher resolution some structures become resolved. One can clearly see
the Local supercluster at the center of the map and that both the GA
and PP may consist of different sub-structures.

The velocity field along the Supergalactic plane is presented in
Fig. 7, showing the existence of two convergence regions which roughly
coincide with the locations of the GA and PP.

\begin{figure}
\setlength{\unitlength}{1cm} \centering
\begin{picture}(7,9)
\put(-3.5, -4){\includegraphics{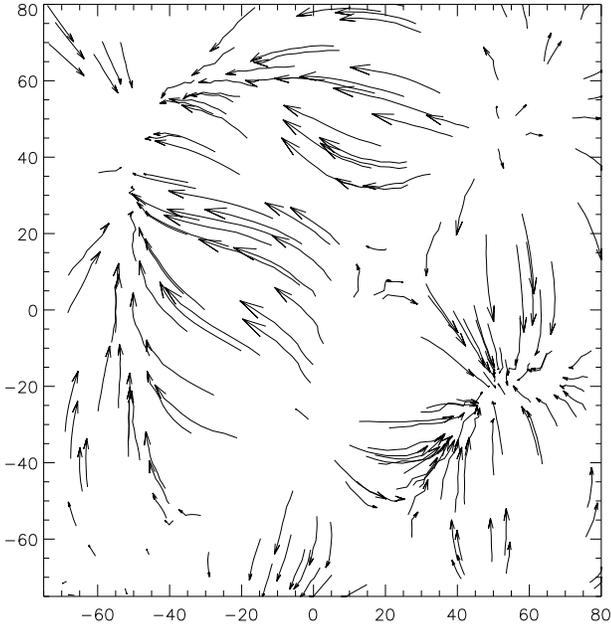}}
\end{picture}
\caption{The G12 reconstructed velocity field in the Supergalactic
plane is displayed as flow lines that start at random
points, continue tangent to the local velocity field, and are of
length proportional to the magnitude of the velocity at the starting
point.}
\label{fig:fig7}
\end{figure}

\subsection{Bulk Velocity}

The velocity field has been fitted using a monopole, dipole (\ie\ bulk
flow) and quadrupole (\ie\ shear) expansion within spheres of radii
ranging from $1000$ to $6000\kms$.  The three Cartesian components of
the bulk velocity (in Supergalactic coordinates) and its absolute value
($V_{B}$) are shown in Fig. 8 as a function of the depth over which the
fit has been done.  The plots present the bulk velocity of the WF field
and of an ensemble of 10 CRs. The plot of the absolute value of the bulk
velocity contains also the mean and standard deviation calculated over
the ensemble of the CRs. Note that the mean $V_{B}$ of the CRs is higher
than its WF value. This result is expected as the WF attenuates the
velocity field with the depth, as the observational errors become more
dominant. 

The amplitude of the bulk flow measured from the reconstructed
three-dimensional velocity field ranges from $V_{B} =300 \pm 70\kms$
for a sphere of $R=20\hmpc$ to $160 \pm 60\kms$ for $R=60\hmpc$.  This
value is in good agreement with that obtained from a direct fit to the
radial peculiar velocities (da Costa \etal\ 2000b).  This result
disagrees with the bulk flow determination from the {\rm Mark III} survey where
the amplitude is roughly twice that of ENEAR with a comparable scatter
(Zaroubi \etal 1999) but comparable to that measured from the SFI
sample.

\begin{figure}
\setlength{\unitlength}{1cm} \centering
\begin{picture}(8,9)
\put(-1., 0){\includegraphics{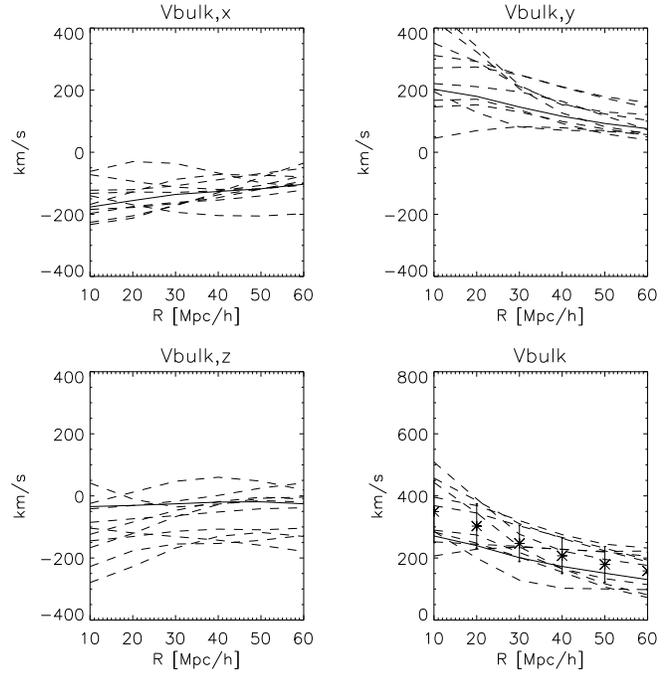}}
\end{picture}
\caption{The bulk velocity fit of the reconstructed velocity field as
a function of depth: The solid line corresponds to the WF field and
the dashed lines correspond to an ensemble of 10 CRs.  The four panels
show the (Super Galactic) x,y and z components and the amplitude of
the bulk velocity.  The bottom-right panel presents also the mean
amplitude taken over the CRs and the error-bars are the standard
deviation around it.  Note that this mean value is expected to be
larger than the amplitude of the WF bulk velocity.}
\label{fig:fig8}
\end{figure}

\subsection{Large Scale Tidal field}

An alternative way of describing the velocity field is to decompose it
into two components, one which is induced by the local mass
distribution and a tidal component due to mass fluctuations external
to the volume considered.  Here we follow the procedure suggested by
Hoffman (1998a,b) and more recently by Hoffman \etal\ (2000).  The key
idea is to solve for the particular solution of the Poisson equation
with respect to the WF density field within a given region and zero
padding outside.  This yields the velocity field induced locally,
hereafter the divergent field.  The tidal field is then obtained by
subtracting the divergent field from the full velocity field. Fig.  9
shows the results of this decomposition applied to the ENEAR survey,
where the local volume is a sphere of $ 80\hmpc$ centered on the Local
Group.  The plots show the full velocity field (upper left panel), the
divergent (upper right panel) and the tidal (lower left panel)
components.  To further understand the nature of the tidal field its
bulk velocity component has been subtracted and the residual is shown
in the lower right panel.  This residual is clearly dominated by a
quadrupole component.  In principle, the analysis of this residual
field can shed light on the exterior mass distribution.

For the ENEAR catalog we find that the local dynamics is hardly affected
by structure on scales larger than its depth. For this sample not only
the bulk velocity at large radii is small but so is the rms value of the
tidal field estimated to be of the order of $60\kms$. This is in marked
contrast to the the results obtaine from the analysis of the {\rm Mark III} survey
which yields a much stronger tidal field, pointing (in the sense of its
quadrupole moment) towards the Shapley concentration. For {\rm Mark III} the tidal
field contributes roughly third of the total bulk velocity ($\sim
200$km/s).

\begin{figure}
\setlength{\unitlength}{1cm} \centering
\begin{picture}(8,10)
\put(-2,-.5){\includegraphics{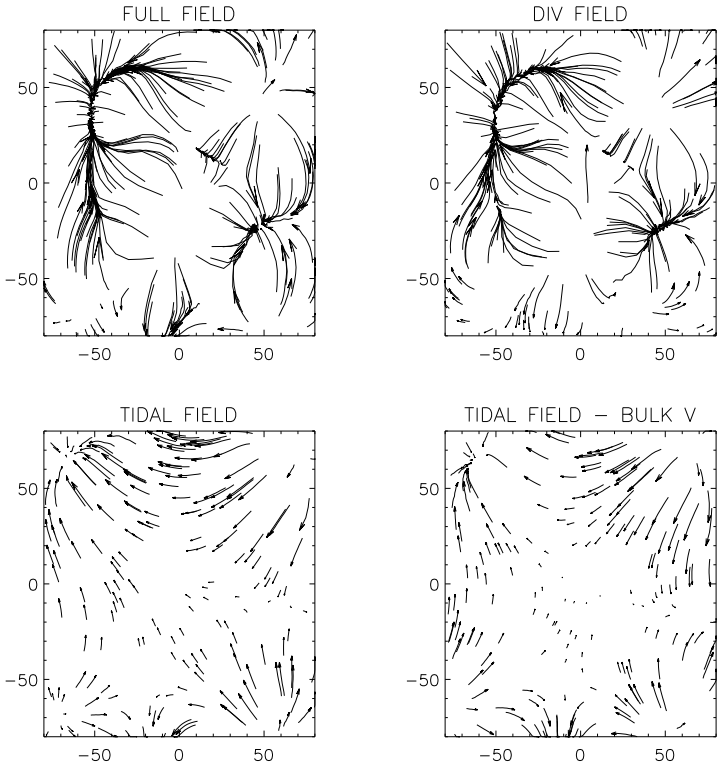}}
\end{picture}
\caption{Tidal field decomposition of The G5 reconstructed velocity
field in the Supergalactic plane is displayed as flow lines. The
top-left panel shows to the full velocity field.}
\label{fig:fig9}
\end{figure}

\section{CONCLUSION}
\label{sec:conclusion}

In the first part of this paper the maximum-likelihood method (Zaroubi
\etal 1997) has been used to measure the mass-density power spectrum
from the newly completed ENEAR early--type redshift-distance
survey. The method assumes that the galaxy peculiar velocities satisfy
Gaussian random statistics and that they are linearly related to the
mass-density field. The initial fluctuation power spectrum is assumed
to be CDM-like, with or without COBE normalizations. In addition the
measured peculiar velocities error are assumed to be proportional to
the distance with some thermal component to account for the nonlinear
evolution of high-density environment in which the early--type
galaxies reside.
 
General results that are valid for all the models used in the
analysis, and are independent of the detailed parameterization and
normalization used in each model, can be summarized as follows.  The
amplitude of the power spectrum at $k=0.1 \ihmpc$ is
$P(k)\Omega^{1.2}= (6.5\pm 3) \times 10^3 (\ihmpc)^3$ yielding
$\eta_8=1.1^{+0.2}_{-0.35}$. For the family of COBE-normalized CDM
models the following range of parameters was considered: $\Omega \le
1$; $0.4 < h <1$; and $n=1$. Within this range we have obtained a
constraint on a combination of the parameters $\Omega$ and $h$ which
can be approximated by $\Omega\approx (0.38\pm 0.08) h^{-1.3}$ for
\lcdm, and $\Omega\approx (0.52\pm 0.083) h^{-0.88}$ for \ocdm. For
$h=0.65$, \lcdm\ yields $\Omega= 0.5-0.8$.  Similar constraints are
obtained from the analysis of the generic $\Gamma$-models, independent
of the COBE normalization. We find that the power spectrum amplitude
and shape parameter are constrained to be $\eta_8=1.0^{+0.3}_{-0.28}$
and $\Gamma \ge 0.18$, with larger values of $\Gamma$ ($>0.4$) being
more probable. We point out that these constraints are consistent with
the results obtained from a similar analysis of the {\rm Mark III} and the SFI
peculiar velocity catalogs. This agreement is encouraging since it
shows that the results are robust and independent of the sample used.

Examination of the $\chi^2/d.o.f.$ for the most likely COBE-normalized
models shows that their values are of the order of 0.93. These values
are about $2\sigma$ away from the preferred value of 1. However, this
should not be too alarming as many of the models within the errors
have $\chi^2/d.o.f \sim 1$. The $\chi^2/d.o.f$ for the best-fit
$\Gamma$-model is 0.99.

As pointed out by previous papers that have analyzed the PS derived
from peculiar velocity data (Zaroubi \etal 1997, Freudling \etal
1999), the constraints on $\eta_8$ and $\Gamma$ are considerably
higher than those obtained from other types of analyzes including
peculiar velocity data (Borgani \etal 1997, 2000), cluster abundances,
and the galaxy power-spectrum (Efstathiou \etal 1992; Sutherland \etal
1999). They are also not consitent with those obtained by combining
the results from high redshift supernovae type Ia (Perlmutter \etal
1999) and the CMB data (Efstathiou \etal 1999) which yields values of
$\Omega \approx 0.25\pm 0.15$ and $\Lambda\approx 0.65 \pm
0.2$. Furthermore, assuming a linear galaxy-mass relation  the value of
$\eta_8$ obtained from the present analysis would imply $\beta=1.0$ or
a $\beta_I \sim 1.4$ (\eg Willmer, da Costa \& Pellegrini 1999;
Sutherland \etal 1999), where the subscript refers to \iras\ galaxies,
at least a factor of 2 larger than those derived from a
velocity-velocity comparison of the \iras\ 1.2 Jy gravity field and the
{\rm Mark III} (Davis \etal 1996), SFI ( da Costa \etal 1998) and ENEAR (Nusser
\etal 2000) all leading to $\beta_I \sim 0.5$. These values are also
consistent with those derived from small-scale velocities (Fisher
\etal 1995).

There are many possible explanations for the above discrepancies. One
possibility is that all other analyses have somehow conspired to
produce consistent results but yet incorrect interpretation. Even
though at first glance this seems unlikely, the recent results from
the CMB ballon experiments Boomerang (Bernardis \etal 2000; Lange
\etal 2000) and MAXIMA (Hanany \etal 2000; Balbi \etal 2000) have
shown that the height of the second peak in the CMB angular
power-spectrum is consistent with higher values of $\Omega$. From
their most likely models these authors derive $\Omega= 0.4-0.8$

It is important to point out that the method is very sensitive to the
assumed error model which can add or supress power. It also implicitly
gives a high weight to nearby galaxies, likely to be slow rotators or
low velocity dispersion systems, for which the measurements and the
distance relations are the least reliable. However, tests have shown
that these effects are not very important for the present data set.
Another potential problem arises due to the rapid decrease of the
weight with distance, the effective volume of the currently available
catalogs is small and the shape of the power spectrum is poorly
constrained, as illustrated by the case of the $\Gamma$-model. All
these factors may impact on the reliability of the constraints
obtained from the PS analysis.

Finally, one or more of the theoretical model ingredients could be
inaccurate, \eg, power spectrum assumed shapes, Gaussianity of the
distribution; or even some inherent bias in the method itself that has
eluded the extensive numerical tests carried out with the data and
mock samples (\eg Freduling \etal 1999). In fact, through an eigenmode
expansion of the {\rm Mark III} and SFI galaxy catalogs, Hoffman and Zaroubi
(2000) have conducted a mode--by--mode goodness--of--fit
analysis. They found that when the surveys are analyzed with their
corresponding CDM most likely models, there is a systematic
inconsistency between the data and the `best-fit' models suggesting
either a generic problem in the peculiar velocity data sets or the
inadequacy of the theoretical or error models. Unfortunately, however,
the analysis has not been able to point to the exact source of
inconsistency.

Finally, in this study we have also performed, given the most probable
power spectrum, a Wiener reconstruction of the density and velocity
fields. The maps shown here have $1200 \kms$ and $900 \kms$ Gaussian
resolution and they are limited to the Supergalactic plane. The main
features shown are similar to the features in the \iras\
reconstruction, corrected for perculiar velocities. The constrained
realizations allow us to estimate the point-by-point uncertainties in
the recovered maps.  In terms of their recovered density fields ENEAR,
SFI and {\rm Mark III} mostly agree. However, they do differ in the velocity
fields. ENEAR shows no significant tidal component which contributes
about half of the {\rm Mark III} local bulk velocity. This tidal field accounts
for the very different bulk velocities obtained from ENEAR and {\rm Mark III},
with SFI situated in between these surveys.  The results suggest that
volumes of $60-80\hmpc$ are essentially at rest relative to the CMB
and that the Local Group motion is primarily due to mass fluctuations
within the volume sampled by the existing catalogs of peculiar
velocity data.

\section{Acknowledgments} We thank Enzo Branchini for providing the PSCz
density field. We acknowledge Avishai Dekel, Enzo Branchini, Tony
Banday, Ravi Sheth, Simon White and Idit Zehavi for stimulating
discussions. SZ gratefully acknowledge the hospitality of Kapteyn
Astronomical Institute -- Groningen.The authors would like to thank
M. Maia, C. Rit\'e and O. Chaves for their contribution over the
years.  MB thanks the Sternwarte M\"unchen, the Technische
Universit\"at M\"unchen, ESO Studentship program, and MPA Garching for
their financial support during different phases of this research.  MVA
is partially supported by CONICET, SecyT and the Antorchas--Andes--
Vitae cooperation. GW is grateful to the Alexander von
Humboldt-Stiftung for making possible a year's stay at the
Ruhr-Universit\"at in Bochum, and to ESO for support for visits to
Garching which greatly aided this project.  Financial support for this
work has been given through, Israel Science Foundation grant 103/98
(YH), FAPERJ (CNAW, MAGM, PSSP), CNPq grants 201036/90.8, 301364/86-9
(CNAW), 301366/86-1 (MAGM); NSF AST 9529098 (CNAW); ESO Visitor grant
(CNAW). PSP and MAGM thank CLAF for financial support and CNPq
fellowships.  The results of this paper are based on observations at
Complejo Astronomico El Leoncito (CASLEO), operated under agreement
between the Consejo Nacional de Investigaciones Cient\'\i ficas de la
Rep\'ublica Argentina and the National Universities of La Plata,
C\'ordoba and San Juan; Cerro Tololo Interamerican Observatory (CTIO),
operated by the National Optical Astronomical Observatories, under
AURA; European Southern Observatory (ESO), partially under the ESO-ON
agreement; Fred Lawrence Whipple Observatory (FLWO); Observat\'orio do
Pico dos Dias, operated by the Laborat\'orio Nacional de Astrof\'\i
sica (LNA); and the MDM Observatory at Kitt Peak 

{}

\end{document}